\documentclass[slac_one]{revtex4}
\usepackage{graphicx}
\usepackage{fancyhdr}
\newcommand{\gtwo}  {$g_{2}$}
\newcommand{\Z}  {$Z$}
\newcommand{\Zg}  {$Z/\gamma^*$}
\newcommand{\W}  {$W$}
\newcommand{\al}  {$a_{L}$}\newcommand{\at}  {$a_{T}$}
\newcommand{\pt}  {$p_{T}$}
\newcommand{\pT}  {$p_{T}$}
\newcommand{\ET}  {$E_{T}$}
\newcommand{\resbos}  {{\sc resbos}}
\newcommand{\pythia}  {{\sc pythia}}
\newcommand{\mcfm}  {{\sc mcfm}}
\newcommand{\alpgen}  {{\sc alpgen}}
\newcommand{\sherpa}  {{\sc sherpa}}

\begin{document}

%Title of paper
\title{\boldmath{\Z} (+jets) at the Tevatron} %% Paper title goes here

% Repeat the \author .. \affiliation  etc. as needed
%
% \affiliation command applies to all authors since the last
% \affiliation command. The \affiliation command should follow the
% other information

\author{G. Hesketh, on behalf of the D0 and CDF collaborations}
\affiliation{Northeastern University, Boston MA, USA}

\begin{abstract}
We report on recent \Z\ (+jets) measurements from the Fermilab Tevatron proton anti-proton collider. 
A new D0 measurement of the transverse momentum of \Z\ bosons yields the best measurement to date of the non-perturbative form factor, \gtwo.
The production of \Z+jets is an major background to many rare signals, and is a vital testing ground for theoretical predictions.
Measurements from CDF and D0 of differential cross sections in \Z+jet production test NLO pQCD, and in the case of D0, the latest tree-level matrix element with matched parton shower calculations.
Improving modelling of this signal will impact results from the Fermilab Tevatron and CERN LHC.
\end{abstract}

\maketitle

\thispagestyle{fancy}

\section{INTRODUCTION} % Section title should be in all capitals.
The production of massive gauge bosons, like the $W$\ and $Z$, are important signals at hadron colliders such as the Fermilab Tevatron and CERN Large Hadron Collider.
The electron and muon decay modes provide distinct experimental signatures, and can be identified with low backgrounds.
Such events can be used as probes of the underlying QCD, to study the production mechanism of heavy bosons, and the production of additional hard partons in association with those bosons.

The large samples of the \W\ boson available at the Tevatron are also yielding the most accurate \W\ boson mass measurements.
The majority of \W\ (and \Z) bosons are produced with little momentum transverse to the beam direction (\pt), mostly recoiling against soft gluon emission. 
Understanding this \pt\ spectrum is important to precision measurements such as the \W\ boson mass, where there is some ambiguity between the boson \pt\ and the missing \pt\ due to the unreconstructed neutrino from the \W\ boson decay.
This \pt\ distribution is best studied in \Z\ boson production, where the \Z\ boson (decaying to electron or muon pairs) is fully reconstructed.
Theoretically, the low boson \pt\ region is modelled by gluon re-summation, such as in the BNLY~\cite{bnly} parameterization, which involves three form factors which must be measured experimentally. 
One of these factors, \gtwo, can be extracted from the shape of the \Z\ boson \pt\ distribution.
The D0 experiment present a new measurement of \gtwo.

Production of hard partons in association with \W\ and \Z\ bosons results in a complex final state, and one that is common to many rare signals, such as top decay, associated production of the Higgs boson, and the production of some super-symmetric particles.
In order to search for and study these rare signals under the huge Standard Model \W\ or \Z+jet background, an accurate model of this background is needed.
The current theoretical predictions have reached next-to-leading (NLO) in perturbative QCD (pQCD) for boson + $\le 2$\ parton production~\cite{mcfm}. 
For event generators, the current best approach is to use tree-level matrix element calculations, with matched parton showering (ME+PS)~\cite{alpgen, sherpa}. 
As these are leading order matrix element calculations, they suffer significant scale uncertainties and must be tuned to reproduce real data.
Both the CDF and D0 experiments present measurements of differential cross sections in \Zg+jet(s) production and test NLO pQCD, and in the case of D0, also test the ME+PS event generators.

\section{TRANSVERSE MOMENTUM OF THE \boldmath{\Z} BOSON AND \boldmath{\gtwo}}

Previously, D0 measured the cross section for \Zg\ production, differential in \Zg\ \pT ~\cite{z_pt_pub}.
The measurement was made using approximately 1.0~fb$^{-1}$\ of integrated luminosity, looking at the $Z\rightarrow ee$\ mode by selecting events containing two electron candidates reconstructed in either the central calorimeter ($|\eta|<1.1$, where $\eta = -ln(tan(\theta/2))$, and $\theta$\ is the polar angle measured with respect to the proton beam direction), or the forward calorimeters ($1.5<|\eta|<3.2$).
Electrons are required to have  \pt$>25$~GeV, and the di-electron mass lie between 70 -- 110~GeV, consistent with the \Z~boson.

\begin{figure*}[t]
\centering
\includegraphics[width=80mm]{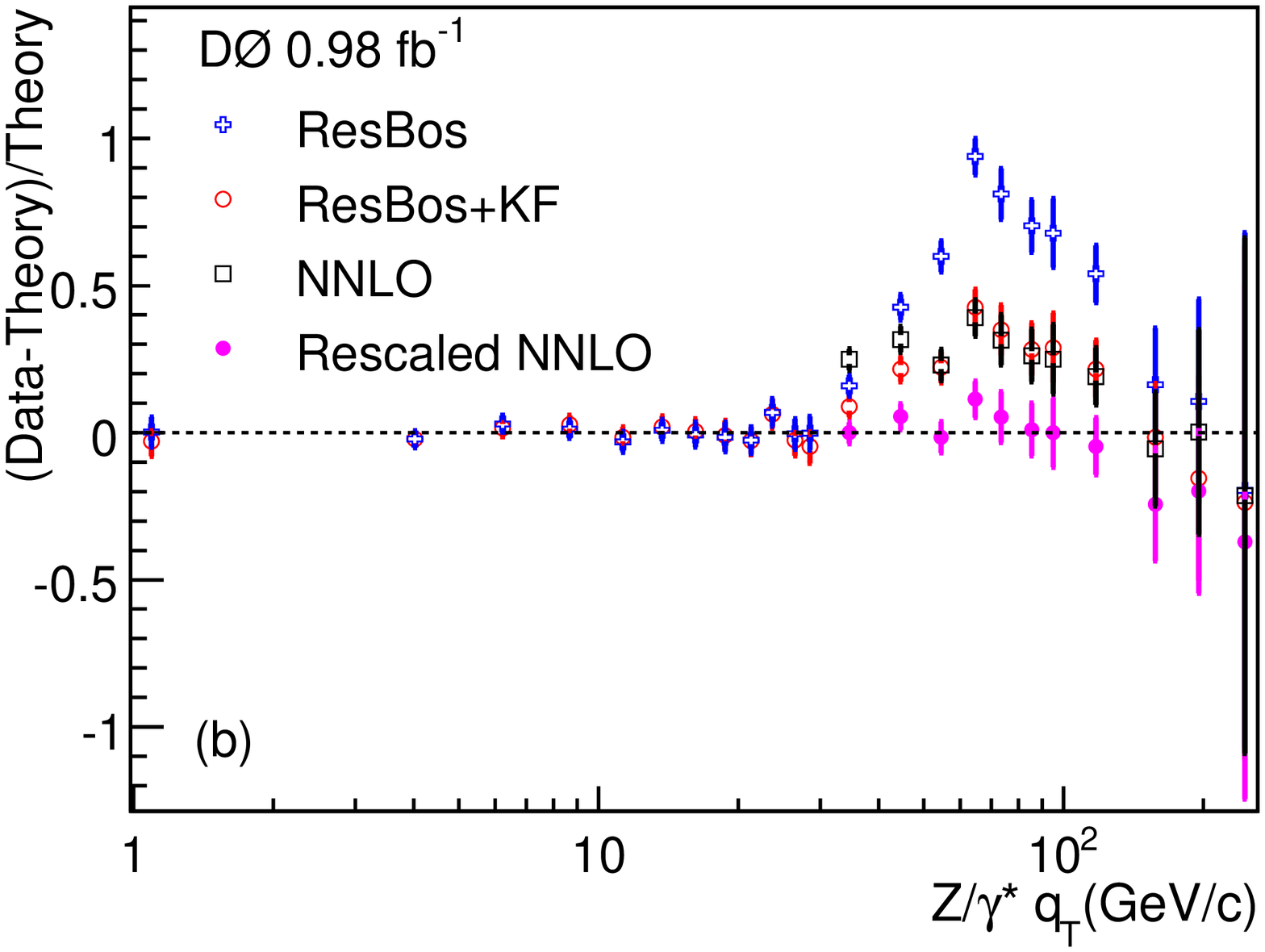}
\includegraphics[width=80mm]{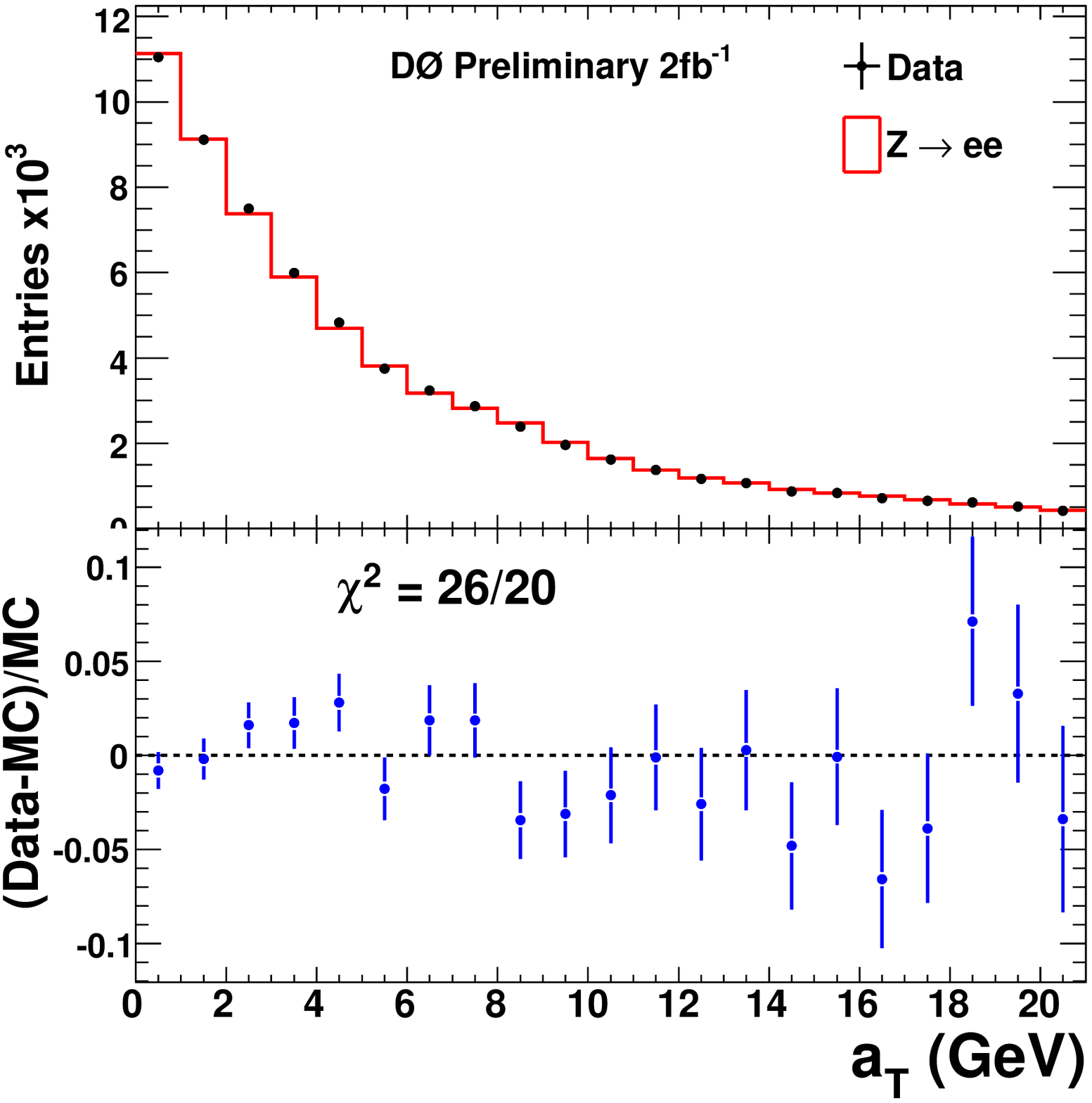}
\caption{Comparing the  corrected \Zg\ \pt\ distribution (left), and \at\ at detector level (right) with theory.} \label{fig:d0_zpt}
\end{figure*}

NNLO predictions~\cite{nnlo_z} are compared to measured differential cross section at high \Zg~\pt, and found to describe the shape well, but need a scale factor of 1.25 to match the normalization (see figure \ref{fig:d0_zpt}).
The prediction of {\sc resbos}~\cite{resbos}, which contains the gluon re-summation parameterization,  describes the data well at low \Zg\ \pt, with an optimal value of $g_{2} = 0.77 \pm 0.06$, limited by experimental resolution.
Bosons produced at high rapidities ($y$) are expected to be sensitive to the ``small-$x$\ broadening'' effect~\cite{smallx3}. 
Studying the \pt\ distribution of \Z\ bosons with $|y|>2$\ shows a slight preference for the prediction without small-$x$\ broadening, but the measurement is limited again by experimental resolution.

To improve these results, D0 have adopted a new variable. 
After reconstructing the leptons from the \Zg\ decay, the thrust axis is calculated, and the \Zg\ \pt\ decomposed into a component parallel to the thrust axis (\al), and a component perpendicular (\at). 
The variable \at\ is largely insensitive to detector resolution, and is used to extract a more precise measurement of \gtwo.
The measurement uses approximately 2.0~fb$^{-1}$\ of integrated luminosity, and combines both electron and muon channels, with muons required to have $|\eta|<2$\ and \pt$>15$~GeV.
The preliminary result presented is not fully corrected for experimental resolution and acceptance, so cannot be directly compared to  \resbos\ to extract \gtwo. 
Instead, \resbos\ samples are generated in with various \gtwo\ values, then samples of \pythia\ with full detector simulation are re-weighted to match the \resbos\ predictions.
These \pythia\ samples are fitted to the data (see Figure \ref{fig:d0_zpt} for one such fit), and extrapolating between them yields the best fit of $g_{2} = 0.63 \pm 0.02 \pm 0.04$, where the first uncertainty is experimental and statistics dominated, and the second uncertainty come from the PDF ({\sc cteq}6.6~\cite{cteq}).
This is of comparable accuracy to the current world average, $g_{2} = 0.68 ^{+0.02}_{-0.01}$, and will improve with increased statistics.

\section{\boldmath{\Z}+JET PRODUCTION}

Both CDF and D0 have studied the production of  \Zg\ in association with jets. 
CDF study the electron \Zg\ decay mode, selecting two electrons with \ET$< 25$, one electron with $|\eta|<1.0$, the other with  $|\eta|<1.0$ or $1.2<|\eta|<2.8$, and a di-electron mass between 66 and 116~GeV.
To reduce backgrounds from jets mis-reconstructed as electrons and from semi-leptonic decays, electrons are required to be isolated from any hadronic activity in the form of charged tracks or energy in calorimeter cells.
D0 look at the muon mode, selecting two muons with opposite charge, \pt$>15$~GeV, $|\eta|<1.7$, again with isolation requirements to reduce backgrounds to negligible levels.
Both experiments use a seeded mid-point cone jet reconstruction algorithm, though with some technical differences~\cite{cdf_jets, dzero_jets}.
CDF use a cone size of 0.7 to reconstruct jets, requiring those jets to have $|\eta|<2.1$\ and \pT$>30$~GeV. 
D0 use 0.5 for the cone size, requiring jets to have \pT$>20$~GeV, and $|y|<2.8$.

The main challenge in extracting differential cross sections from \Zg+jet distributions is the understanding of the experimental resolution, particularly for jets. 
Both experiments have put a great deal of effort into understanding the jet energy scale and resolution, but these still dominate the systematic uncertainties on these measurements.
The experiments use different techniques  to correct for the effects of resolution on the distributions, but in both cases the resulting method systematics are small.

\begin{figure*}[t]
\centering
\includegraphics[width=80mm]{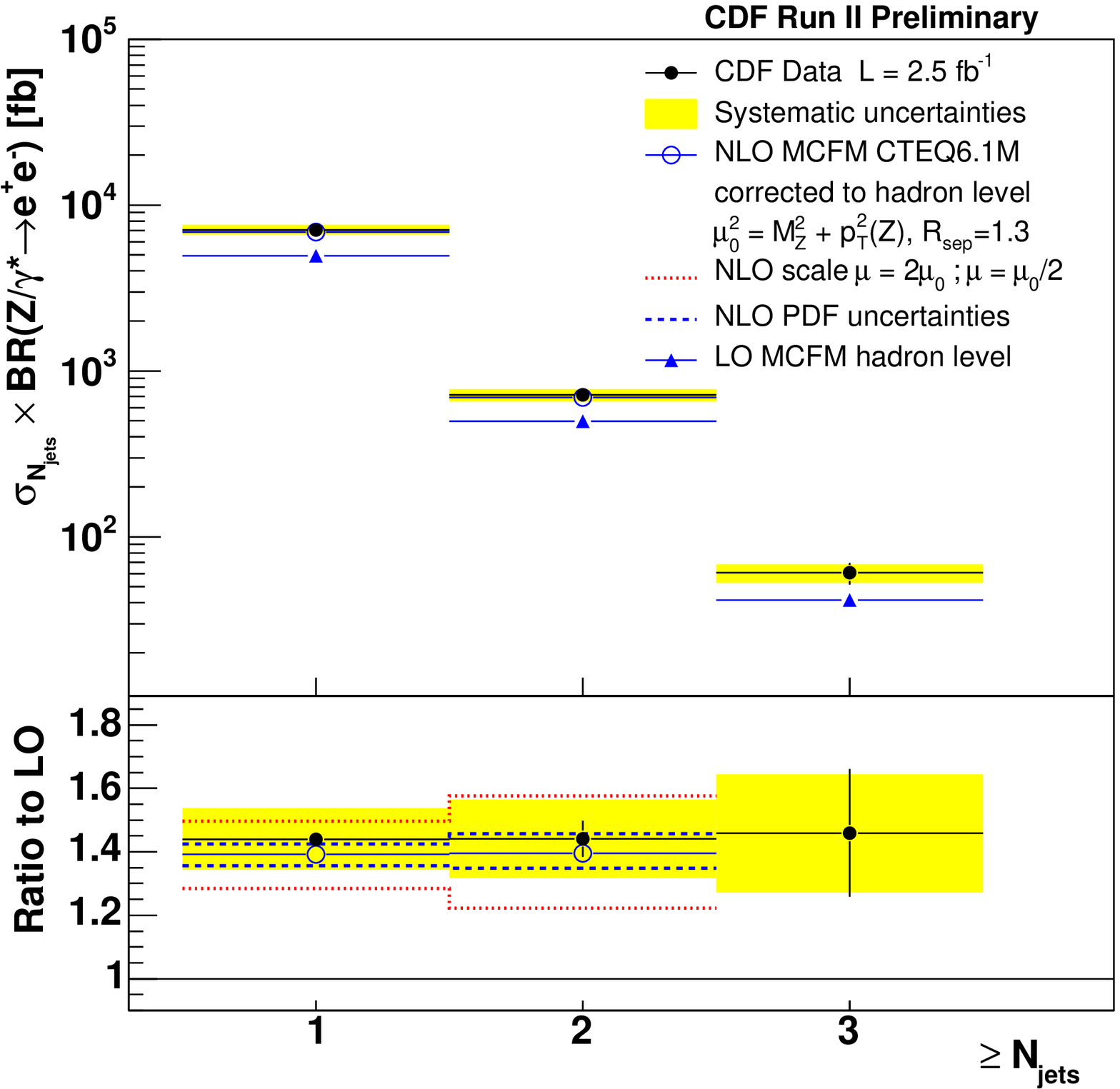}
\includegraphics[width=80mm]{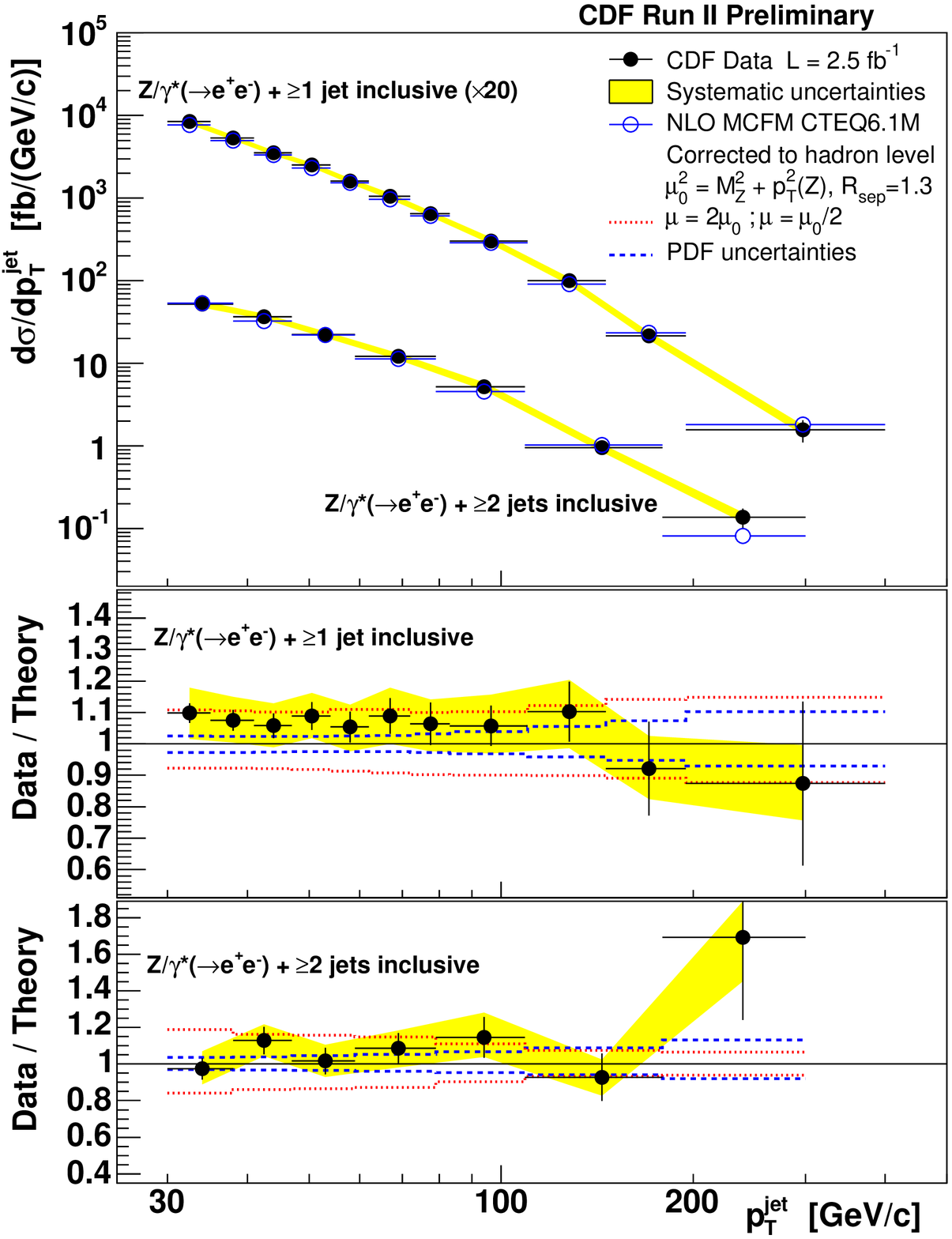}
\caption{Jet yields (left) and \pt\ distributions (right) in \Zg\ events from CDF.} \label{fig:cdf_yields}
\end{figure*}

CDF present an updated result using more integrated luminosity (2.5~fb$^{-1}$) than an earlier publication~\cite{cdf_zjets}, and have measured the yields for one, two and three jets in \Zg\ events, and cross sections differential in inclusive jet pT and inclusive jet rapidity, for events containing at least one and at least two jets.
The jet yields  and the \pt\ results are shown in Figure \ref{fig:cdf_yields}.
Leading order (LO) and NLO pQCD predictions from \mcfm\ are compared to the measurements, after applying non-perturbative corrections derived from an event sample generated with \pythia~\cite{pythia}, which take the parton-level NLO pQCD prediction to the particle level.
The resulting distributions agree with the data.

\begin{figure*}[t]
\centering
\includegraphics[width=80mm]{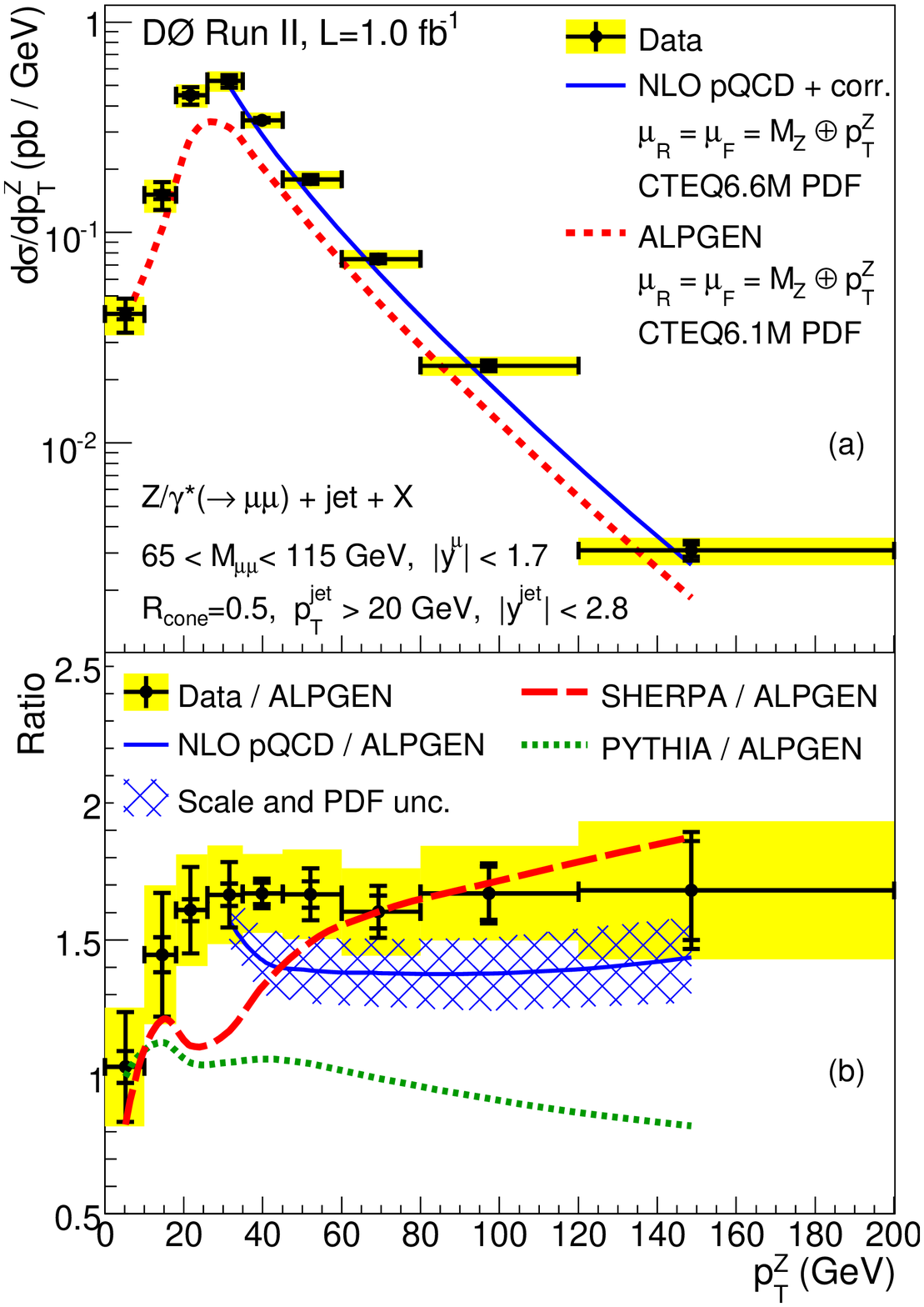}
\includegraphics[width=80mm]{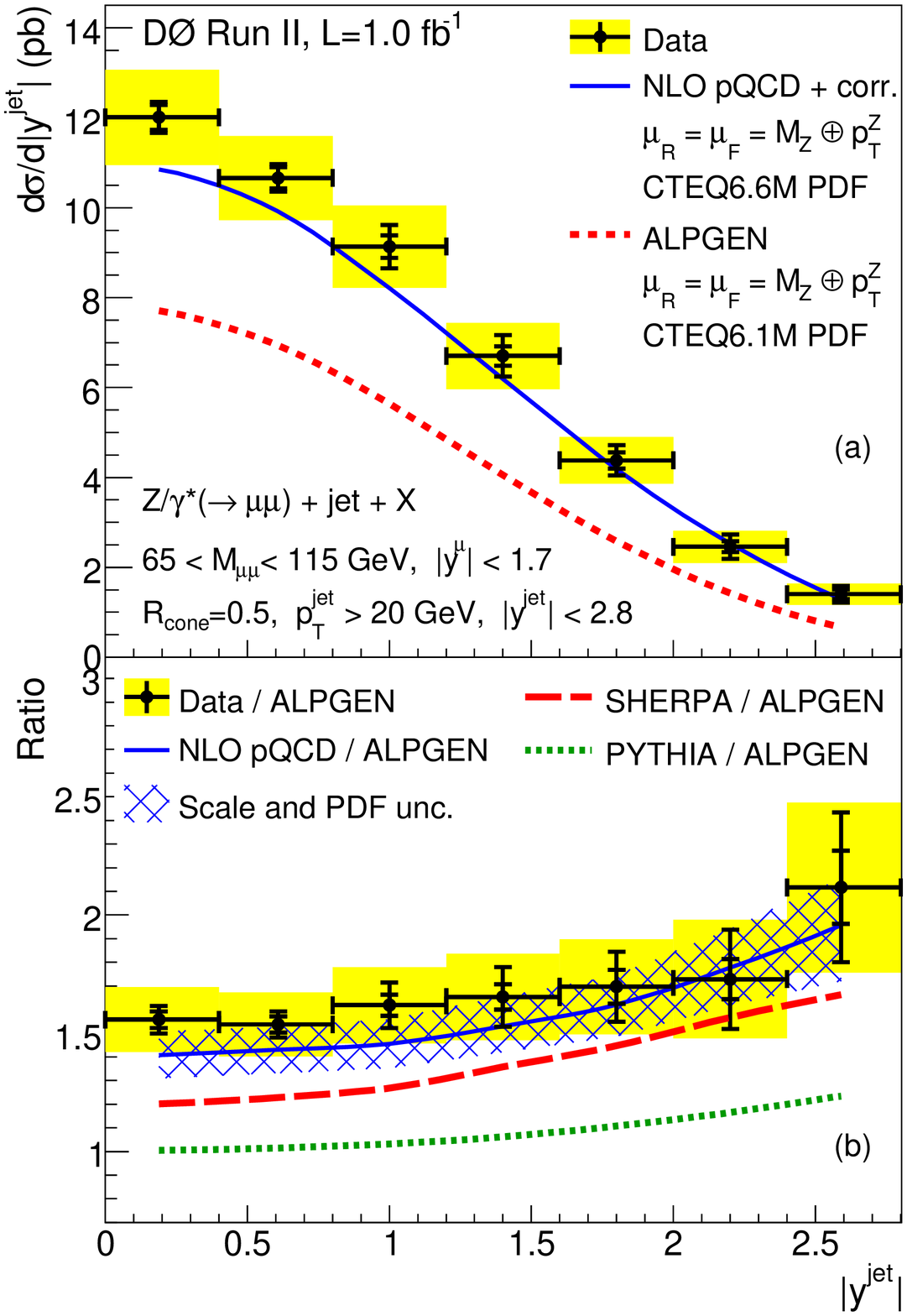}
\caption{\Zg\ \pt\ in events with at least one jet (left), and jet rapidity in \Zg\ events (right) from D0.} \label{fig:d0_jets}
\end{figure*}

D0 have measured the cross section for \Zg+jet+X production, differential in the leading jet \pt\ and rapidity, and the \Zg\ \pt\ and rapidity~\cite{dzero_zjets}.
Again, NLO pQCD predictions with non-perturbative corrections applied are compared to the measured cross sections and show good agreement, except at low \Zg\ \pt\ where non-perturbative processes dominate and the prediction is not shown. 
Additionally, the predictions from three event generators are compared:
i) \alpgen, a ME+PS generator, using \pythia\ for the showering;
ii) \sherpa, also a ME+PS generator;
iii) \pythia, with all jets coming from the parton shower.
All generators show significant normalization differences to the data, and the shapes are best described by \alpgen. 
However, the low \Zg\ \pt\ region, which is particularly sensitive to the description of jets coming from the underlying event, is not well described, and the jet rapidity distribution predicted by \alpgen\ appears too narrow.
These distributions can be seen in figure \ref{fig:d0_jets}

With more luminosity these results can be extended, placing tighter constraints on the high \pt\ tails, and on higher jet multiplicities. 
These are important measurements, testing NLO pQCD, and the modelling of these complex final states by event generators.
Understanding these processes is vital to the sensitivity to new physics at the Tevatron and LHC.

\end{document}